# Systematic Approach on Differences in Avian Viral Proteins Based on Carbon Composition

BABY JERALD A. #1

#  Research Associate, RIIC

Bioinformatics Research Group

Dayananda Sagar Institutions

[2] babyjerald@gmail.com

T.R. GOPALAKRISHNAN NAIR*2

ARAMCO Endowed Chair – Technology, PMU, KSA.

* Vice President – RIIC,

Dayananda Sagar Institutions

[2] trgnair@gmail.com

*Abstract:* The distribution of amino acid along the protein sequences plays an imperative role in facilitating different biological functions. Currently, there is insufficient scientific data, which represents the arrangement of amino acid in the proteins based on atomic composition. Our deep observational and analytical studies indicate that the distribution of carbon in the protein sequence can bring differences in the function of proteins. Until now, it is believed that carbon content elicits the hydrophobic interactions in proteins. This distinct feature classifies normal proteins and viral proteins based on the carbon content. One of the objectives of this investigation is to show the significance of carbon composition in protein structure evaluation. Since, the level of perceived benefit is likely to be high in the field of proteomics for structural analysis, the position of this paper is to prioritize only the avian viral protein sequences based on carbon content when compared with the normal protein sequences. Here, we present the evaluation of carbon along the avian protein sequence in order to show the differences in the distribution of carbon using the software technology. The evaluation results provide a deep view in to the molecular structure of avian viral protein, which can further enable the progress of research in the proteomics domain based on carbon classification. This systematic approach in computing carbon level of avian viruses aim towards the benefit of individual, regional and global health in addition to opening of further effective research on avian viral protein based on the aforementioned results.

1. Introduction:

Most of the studies have unraveled many aspects of protein structural composition where it is indicated that Carbon, Nitrogen, Phosphorus are the main elements involved in the formation of mRNAs and proteins [1]. In contrast, the level of carbon distribution along the protein sequence varies between normal and viral proteins, though it is organized with same number of amino acids but it makes difference with the arrangement of amino acid. Ultimately ends in different biological function [2]. In a situation where influenza pandemic is considered imminent or underway, a profound investigation in to the avian viral protein structure is fundamental to demonstrate atomic composition for better understanding the differences of protein structure for further research. In this case study, some of the avian viral proteins from different regions is preferred for the analysis based on carbon content. Although the stability and folding nature of protein is accompanied with the carbon content, it is been reported that 31.44% carbon induces the stable nature of proteins [3]. Despite the fact, while come across the preventive measures to counter arrest the spread of seasonal human influenza, vaccination is the only consort to minimize the rate of infection [4]. From the observational studies, it is found that viral RNA mutates and reassorts and finally develops resistance against vaccines [5]. Henceforth, there arises the problem of vaccine efficacy. In such scenarios, the complete viral protein structure evaluation at the atomic level will help to provoke the possibility in bringing down the mortality rate by further effective research in to the molecular level and can address new strategies in controlling the spread of disease. Moreover what could be the difference ratio between the species based on carbon content? What is the carbon composition of avian viral protein? Will that be stable with 31.44% carbon? To address these issues, the systematic approach is been developed to identify differences in avian viral protein sequences.

2. Background:

As outlined in the preceding reports on carbon content, it is evident that the distribution of carbon plays an important role in gene expression and protein stability. Further studies illustrates that the availability of carbon in the genes influences evolution by gene duplication [6]. In contrast to perception, depending on biosynthetic pathways of proteins, each protein found to have assigned value of energetic cost based on the arrangement of amino acids. The differences in the cost of amino acids between different organisms can influence evolutionary implications on different species [7]. Though variations observed in many of the atomic contents of protein, larger variation observed in the carbon atom distribution along the sequence [8].

Upon consideration, the structural properties of influenza are essential for this case study. The avian viruses attack the upper respiratory tract and increase the risk factors for humans to become infected. Avian viruses usually get transmitted through aerosols or infected birds [9]. Once entering in to the respiratory tract of the host, it replicates inside the respiratory tract and damages the host cell [10]. The interest of this systematic approach mainly projects on avian viral envelope proteins as there is huge demand in structure evaluation studies. It is been observed that there are two envelope glyco proteins named Hemagglutinin(HA) and Neuraminidase(NA) based on which the influenza virus is classified in to Influenza A and B viruses. Both the viruses possess different antigenic characteristics [11]. The protein HA mediates binding of virus to the host cell and facilitates the entry of viral RNA in to the host cell and the protein NA is involved in the release of progeny virus from the infected cell [12]. The longevity of viral infection depend mainly on the surface antigens and it is likely found to have subtypes ranging from H1 to H16 and N1 to N9 [13]. Currently avian viruses acquires the capacity for sustained person to person transmission and pretense a high risk to human being with increased mortality rate [14].

Efforts are under way to improve the qualitative study on elemental composition of nucleic acids and proteins of avian viruses as there is lack of information. The available scientific evidence says that the amino acids coded by the nucleic acids are correlated with its elemental composition [15]. Earlier analytical studies on chemical properties of amino acids and their anticodons related to the evolution of genetic code is available which demonstrates the correlation between the anticodons and their atomic composition [16, 17 and 18]. Further studies reveal that the carbon content in proteins is correlated with the genome base composition where it is observed that high GC genome content is related to low carbon content of proteins [19]. However the variations in the carbon content may or may not induce biochemical reactions. This work projects in to the differences in the carbon content of avian viral protein hemagglutinin just to show whether the carbon content is high or low and also the possibility of hydrophobic interaction between the molecules based on the carbon distribution along the protein sequence.

3. Methodology:

The validation of proteome carbon has gained its importance in recent research as it plays main role in maintaining the stability of proteins. The purpose of this work is to identify the carbon composition of avian viral protein sequences from different parts of Asia to analyze the difference in the carbon content. How far the distribution of carbon affects the structure of avian viral proteins? and what could be the significant effect based on high or low carbon content? For that reason, the application of software program CARBANA was used for this investigation.

The 6 avian viral protein sequences were retrieved from the NCBI-Influenza virus resource (www.ncbi.nlm.nih.gov/genomes/FLU/FLU.html). Awaiting biological researches in this field depend mostly on this online available database and have been used here for the selection of viral strains for the analysis (Table 1).

CARBANA was accessed through www.rajasekaran.net.in/tools/carbana.html for computing carbon composition of the selected viral strains. The fasta sequence of all the viral proteins was downloaded from the Influenza Virus Resource for the computation and program was allowed to run with the limited window size 700 as the amino acid number of all the virus strain is strictly within the windows size limit. The computed carbon properties of the given sequences were observed and viewed graphically (Fig 1).

**Table 1:** Selected avian viral strains for the carbon analysis.

| S.No | Strain Name | Influenza A Subtypes | Genbank ID | Target protein |
|---|---|---|---|---|
| 1. | A/Korean native chicken/Korea/K040110/2010 | H9N2 | AEF10493.1 | HA |
| 2. | A/Mallard duck/Korea/W401/2011 | H5N1 | AEJ90156.1 | HA |
| 3. | A/Muscovy duck/Fujian/FZ01/2008 | H6N6 | ADF29678.1 | HA |
| 4. | A/chicken/Anhui/A28/2011 | H9N2 | AEQ72917.1 | HA |
| 5. | A/chicken/Bangladesh/11rs1984-2/2011 | H5N1 | AEQ50029.1 | HA |
| 6. | A/duck/Anhui/B25/2010 | H9N2 | AEQ73377.1 | HA |

## 4. Results and Discussion:

The evaluation of carbon composition of avian viral proteins could have a significant impact in further research implications towards control measures of Influenza pandemics. In contrast to perception, it is easy to predict the stability of proteins with the carbon content. As discussed earlier, the folding of proteins depend on the carbon content and ultimately results in some biological function [20]. The criterion is to analyze whether the avian viruses possess same carbon content or does it shows any differences. Our analysis illustrates the distribution of carbon and the differences based on the standard carbon percentage 31.44% to show whether the selected group of proteins have significantly different elemental composition using software technology. Fig 1 illustrates the carbon % in 6 different influenza viruses graphically. From the observed results, each protein found to have high and low carbon contents which are illustrated in the table below.

**Table 2:** Distribution of carbon observed in 6 avian viral protein sequences with its amino acid number.

| S.No | Graphical figure (From fig 1) | Aminoacid Number | Highest carbon content | Lowest Carbon Content |
|---|---|---|---|---|
| 1. | 1a | 35, 36,40,44,45,49,53,54,57, 58, 62, 66, 67, 71. | Nil | 26.29 |
|  |  | 31 and 32 | 43.82 | Nil |
| 2. | 1b | 438 | Nil | 27.09 |
|  |  | 78 | 46.22 | Nil |
| 3. | 1c | 51, 395, 396 478, 522, 524, 569, 570, 573, 334, 397 | Nil | 28.69 |
|  |  | 73 | 45.82 |  |
| 4. | 1d | 374, 375 and 420 | Nil | 28.29 |
|  |  | 222 and 223 | 36.65 | Nil |
| 5. | 1e | 335 | Nil | 27.49 |
|  |  | 190,468,470 and 474 | 36.25 | Nil |
| 6. | 1f | 339, 340, 341, 378, 467 and 507 | Nil | 28.29 |
|  |  | 220 | 37.45 | Nil |

Table 2 characterize the carbon content along the sequence corresponding to the position of aminoacid. The lowest carbon content in all the viral proteins indicates that there is no hydrophobic interaction and may not involve in any biochemical reactions. In contrast, the highest carbon content in the proteins may instigate or restrain biological functions. Except some part of the aminoacids, all the other aminoacids possess high as well as low carbon content which makes difference when compared to the normal stable proteins.

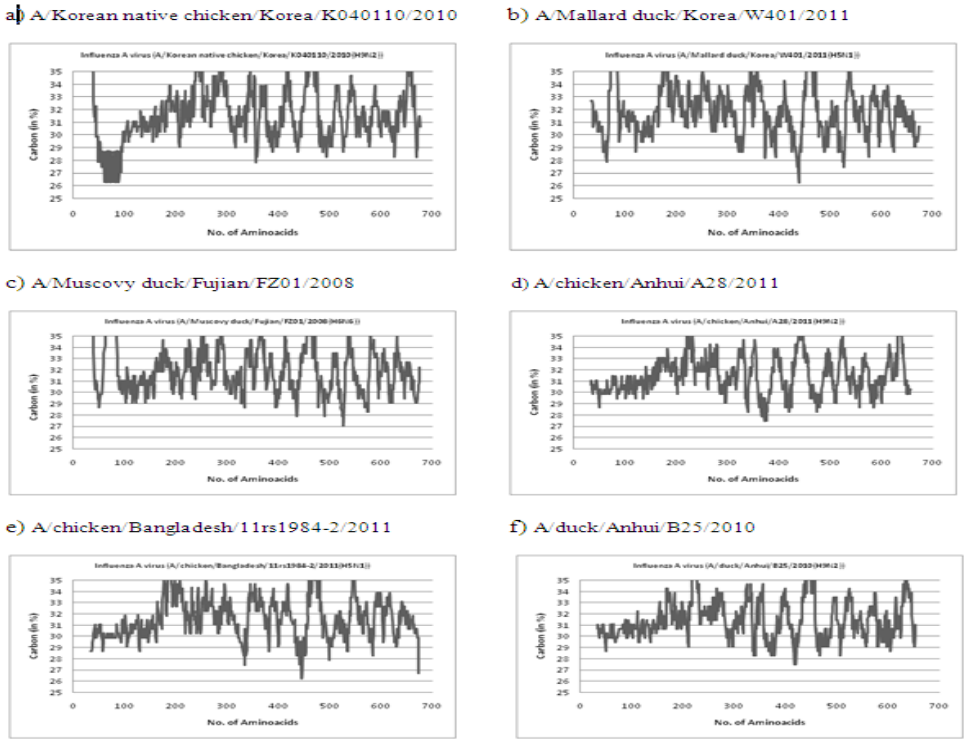

Figure 1: Carbon composition of avian viral protein HA from different regions.

## 5. Conclusion:

Taken together the carbon distribution and the amino acid number, our analyses help explain how the carbon content affects the nature of proteins. In general, it is observed that the avian viral proteins have high carbon content with which it differs from the other normal proteins. As it has got high carbon content, it will tend to have high amino acid cost and consequently plays vital role in evolution by gene duplication. So far the investigation of carbon content confers that 31.44% is needed for stable nature of normal proteins for non covalent interactions with other biomolecules. The availability of high carbon content in avian viral proteins may be encountered as disordered protein which may initiate or suppress biological functions. As the proteins involved in this approach are virulent in nature and spread diseases among population, the aforementioned report on carbon composition would help prevail over the difficulties by further effective research targeting transformation of disorder to ordered proteins and vice versa for the betterment of public health.